# Performance Study of Various Relay Nodes in 5G Wireless Networks

Jianghong Luo, Ashwin Sampath, Navid Abedini, Tao Luo
Qualcomm Technologies, Inc.

*Abstract*—This paper studies performance of various types of relay nodes in a 5G wireless network: conventional amplify-forward repeaters, (semi-)smart/smart amplify-forward repeaters with different levels of side information, and half-duplex/full-duplex decode-forward relay nodes with and without spatial reuse. End-to-end effective signal to interference and noise ratios (SINRs) and achievable rates are derived for these different types of relay nodes. Performance and complexity tradeoffs are discussed with a simulation over a Manhattan topology setting. Over-the-air (OTA) test results corroborates the findings in this paper.

## I. INTRODUCTION

The 5th Generation (5G) of mobile networks promises to provide high data rate services due to availability of large spectrum in the very high frequencies and adoption of large-scale Multiple Input and Multiple Output (MIMO) antenna systems. However, high pathloss and blockage sensitivity at high-frequency bands are obstacles for broader coverage. For example, the coverage of a 5G base station at a Frequency Range 2 band (FR2=24.25GHz-52.6GHz) may be limited to around few hundreds of meters, while the coverage of a typical Frequency Range 1 band (FR1<7.2GHz) macro-cell can be much larger. Signal blockage, at higher bands, will further reduce an effective coverage area of a base station. Therefore, a cost-effective way, e.g. using wireless relay nodes to extend coverage for network densification, is attractive for 5G networks, especially at FR2 and higher bands. The 3rd Generation Partnership Project (3GPP) has had various study and work items in recent years to provide support of different types of relay nodes in a 5G network:

- Integrated access and backhauling (IAB), a Layer-2-based decode-and-forward relay solution, was first introduced in Release-16 [1] and further enhanced in Release-17 and 18.

- Conventional repeaters which simply amplify-and-forward received signal were specified in Release-17, with the corresponding radio frequency requirements in [2].

- Smart repeaters (also called network-controlled repeaters) that perform amplify-and-forwarding operation with side control information provided by the network were studied and specified in Release-18 [3].

- The UE-to-Network Layer 2/Layer 3 relay via side-link was introduced in Release-17 [4] for proximity-based services in 5G systems.

These different types of repeater/relay nodes have different implementation complexities and performance tradeoffs. A number of studies on relay nodes can be found in the literature [5-9], where theoretical performance and bounds on amplify-forward and decode-forward relay operation were derived. But in these studies, some practical constraints for amplify-forward operation, e.g. the constraint on the maximum amplification gain for stability, and noise figure difference between amplify-forward and decode-forward operation, were not captured, and in addition, the performance difference between conventional amplify-forward repeater versus smart amplify-forward repeater as well as impact of different levels of side control information to smart repeater have not been studied. For decode-forward relay nodes, which can be half-duplex or full-duplex, the impact of spatial reuse in case of multi-user scheduling has not been studied before. In this paper, we present an in-depth analysis of performance tradeoffs for different types of repeater/relay nodes that address these points.

An outline of this paper is as follows. Section II describes the system model for different types of repeater/relay nodes and derives end-to-end Signal to Interference and Noise Ratio (SINR) and achievable rates. Section III presents simulation results with mixed base stations and relay nodes in a Manhattan grid deployment setting. Section IV provides main findings from OTA tests conducted in indoor and outdoor environments. Conclusions are drawn in Section V.

## II. SYSTEM MODEL

### A. System Model for Decode-forward Relay node

System model for a decode-forward relay node with downlink (DL) operation is shown in Figure 1, where

- S: signal with unit power.
- $P_{T1}$: transmission power of gNB.
- $P_{T2,max}$: maximum transmission power of relay node.
- $h_1, h_2$: channel states including array beam forming gains for backhaul (BH) and access (AC) links respectively.
- $n_1, n_2$: interference and noise with Gaussian distribution with zero mean and variance $\sigma_1^2, \sigma_2^2$ for BH and AC links respectively.
- $SINR_{BH}, SINR_{AC}$: SINRs for BH and AC links respectively.
- $\beta_{BH}, \beta_{AC}$: fraction of time-domain resources allocated for BH and AC links respectively.

Based on this system model, we have

$$y = h_1\sqrt{P_{T1}}s + n_1,$$

$$z = h_2\sqrt{P_{T2,max}}s + n_2,$$

and the BH and AC SINRs can be calculated accordingly,

$$SINR_{BH} = \frac{P_{T1}\cdot|h_1|^2}{\sigma_1^2}, \quad SINR_{AC} = \frac{P_{T2,max}\cdot|h_2|^2}{\sigma_2^2}$$

The decode-forward relay node can operate based on one of the following two modes:

- *Full-duplex decode-forward mode (FDDF):* in this mode, BH and AC links can operate at the same time with $0 < \beta_{BH} \leq 1$, $0 < \beta_{AC} \leq 1$. Let $C(SINR)$ denote the capacity achieved at SINR. It can be shown that the maximum achievable rate is:
$$C(SINR_{FDDF}) = min(C(SINR_{BH}), C(SINR_{AC})) \quad (1)$$
- *Half-duplex decode-forward mode (HDDF):* in this mode, BH and AC links are time-division-multiplexed with $\beta_{BH} + \beta_{AC} \leq 1$. It can be shown that the optimum resource allocation is achieved when $\frac{\beta_{BH}}{\beta_{AC}} = \frac{C(SINR_{AC})}{C(SINR_{BH})}$, and the resulting achievable rate is given by:

$$C(SINR_{HDDF}) = \frac{1}{\frac{1}{C(SINR_{BH})} + \frac{1}{C(SINR_{AC})}} = \alpha \cdot C(SINR_{FDDF}) \quad (2)$$

where $\alpha \in (0.5, 1)$ and depends on the relative values of $C(SINR_{BH})$ and $C(SINR_{AC})$. If they are the equal, $\alpha = 0.5$; if one value is far larger than the other value, $\alpha \approx 1$. The corresponding end-to-end effective SINR can be obtained by using the inverse function of capacity.

The system model for decode-forward relay node with uplink (UL) operation is similar to DL, except that the 1st hop is AC and the 2nd hop is BH and resulting achievable rate can also be represented by (1) and (2).

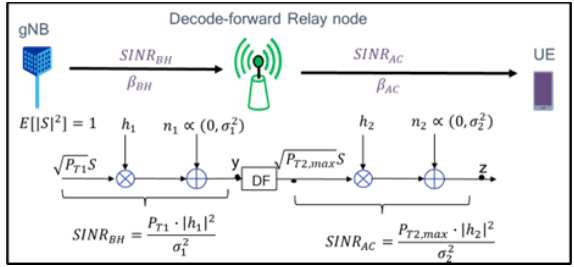

Figure 1: System model for decode-forward relay node in DL

### B. System Model for Amplify-forward Repeater

For the amplify-forward repeater, a unified system model is established for both conventional and smart repeaters with DL operation, as shown in Figure 2.

Here, the amplify-forward repeater is characterized by an amplification gain $G$ and following parameters:

- $G_{max}$: max amplification gain. The repeater adjusts the amplification gain $G$ to achieve a target transmission power $P_{T2,max}$, unless it is limited by $G_{max}$.

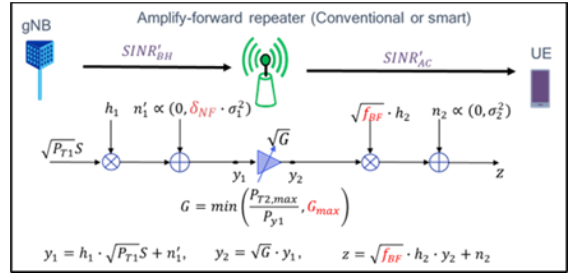

Figure 2: System model for amplify-forward repeater in DL

- $\delta_{NF}$: Noise-figure relative to decode-forward relay node, which captures the noise-figure difference between the RF chains of amplify-forward and decode-forward node depending on the actual implementation. In later simulation, we use $\delta_{NF} = 1dB$.
- $f_{BF} \leq 1$: beamforming loss factor relative to max array gain over AC link:

$$f_{BF} := \frac{actual\ array\ gain}{max\ array\ gain} \leq 1.$$

For a conventional repeater, without knowledge of UE's direction, a fixed broad beam is typically used for AC link to cover all possible directions, and thus the beamforming loss $f_{BF} < 1$; while for a smart repeater with knowledge of UE's direction (provided by the network), a narrow beam can be formed toward UE without any beamforming loss, i.e. $f_{BF} = 1$. Note that this beamforming loss factor is only for AC link. For BH link, it is assumed that the relay node is stationary, and a narrow beam can always be formed over the BH link during deployment.

It can be shown that the final received signal $z$ at the UE side is represented as:

$$z = \sqrt{f_{BF}} \cdot h_2 \cdot \sqrt{G} \cdot \left(h_1 \cdot \sqrt{P_{T1}}S + n_1'\right) + n_2.$$

In order to compare the performance of amplify-forward repeater with decode-forward relay node, the above system model is normalized and rewritten, equivalently, as:

$$\tilde{z} = S + \frac{1}{\sqrt{SINR_{BH}/\delta_{NF}}} \cdot \widetilde{n_1} + \frac{1}{\sqrt{f_{BF}\cdot f_P\cdot f_n\cdot SINR_{AC}}} \cdot \widetilde{n_2},$$

where $(\widetilde{n_1}, \widetilde{n_2})$ are normalized interference and noise with unit variance, and $(SINR_{BH}, SINR_{AC})$ are SINRs of BH and AC links for decode-forward relay node as shown in Section II.A. Parameters $f_P$ and $f_n$ are defined below:

- $f_P \leq 1$: power loss factor resulting from having finite gain in the repeater that sometimes prevents achieving the maximum target transmission power $P_{T2,max}$. The power loss factor depends on the repeater's received power $P_{y1} = \sigma_1^2 \cdot (SINR_{BH} + \delta_{NF})$. If $P_{y1} \geq \frac{P_{T2,max}}{G_{max}}$, there is no loss $f_P = 1$; otherwise, $f_P < 1$.

$$f_P := min\left(1, \frac{P_{y1}\cdot G_{max}}{P_{T2,max}}\right) = min\left(1, \frac{\sigma_1^2\cdot(SINR_{BH}+\delta_{NF})\cdot G_{max}}{P_{T2,max}}\right).$$

- $f_n \leq 1$ : Noise-forwarding loss factor. Note that the transmission power of amplify-forward repeater includes both signal part and noise part. This loss factor captures the ratio of signal power over total transmission power of relay node.

$$f_n \coloneqq \frac{|h_1|^2 P_{T1}}{P_{y1}} = \frac{SINR_{BH}}{\delta_{NF} + SINR_{BH}} \leq 1.$$

Let $SINR'_{BH} = SINR_{BH}/\delta_{NF}$, $SINR'_{AC} = f_{BF} \cdot f_P \cdot f_n \cdot SINR_{AC}$. It can be shown that the end-to-end effective DL SINR for amplify-forward node is given by:

$$SINR_{AF,DL} = \frac{1}{\frac{1}{SINR'_{BH}} + \frac{1}{SINR'_{AC}}}, \quad (3)$$

and the resulting achievable rate is $C(SINR_{AF,DL})$.

Similarly, the system model for amplify-forward repeater node with UL operation can be seen in Figure 3.

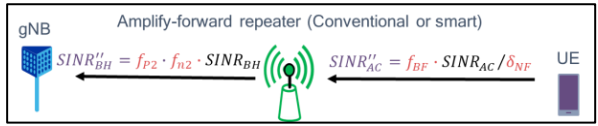

*Figure 3: System model for amplify-forward repeater in UL*

For UL, the 1st hop is AC link and the 2nd hop is BH link. The noise figure $\delta_{NF}$ is associated with 1st hop, loss factors ($f_{P2}, f_{n2}$) are associated with 2nd hop, and the beamforming loss factor $f_{BF}$ is always associated with AC link. The resulting UL effective end-to-end SINR is:

$$SINR_{AF,UL} = \frac{1}{\frac{1}{SINR''_{BH}} + \frac{1}{SINR''_{AC}}}$$

wherein $SINR''_{BH} = f_{P2} \cdot f_{n2} \cdot SINR_{BH}$, and $SINR''_{AC} = f_{BF} \cdot SINR_{AC}/\delta_{NF}$.

The effective SINR and achievable rate calculation can be extended to a general N-hop network, shown in Figure 4, as follows:

- the end-to-end achievable rate for full-duplex decode-forward relay nodes with N hops is $NhopC_{FDDF} = min(C_1, C_2, \ldots C_N)$, with $C_i$ being the capacity of the i-th hop.
- the end-to-end achievable rate for half-duplex decode-forward relay nodes with N hops is $NhopC_{HDDF} = min(C_{HDDF,2}, C_{HDDF,3}, \ldots C_{HDDF,N})$, with $C_{HDDF,i} = \frac{1}{\frac{1}{C_i} + \frac{1}{C_{i-1}}}$ being the capacity of the i-th two-hop system with the optimum resource allocation as shown in (2).
- the end-to-end SINR for amplify-forward repeaters with N hops can be expressed recursively as $effSINR_N = \frac{1}{\frac{1}{effSINR_{N-1}} + \frac{1}{SINR'_N}}$.

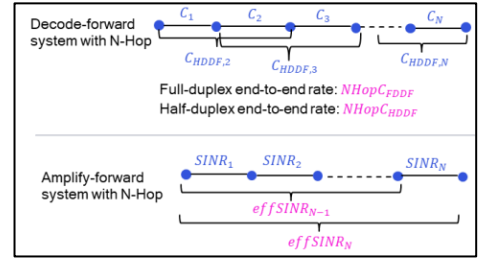

*Figure 4: Extension to N-Hop network*

### C. Benefits of Side Control for Smart Repeaters

Comparing with a conventional repeater, a smart amplify-forward repeater has side information provided by the network to improve the performance over the conventional repeater. In this section, we discuss the impact of two types of side information for smart repeater: *TDD DL/UL configuration* information and the *scheduled beam information* associated with scheduled UE for AC link.

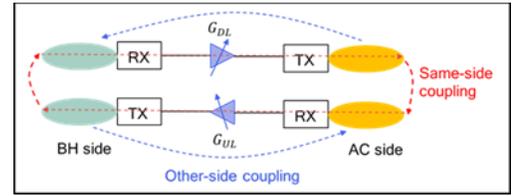

*Figure 5: Stability concerns for amplify-forward repeater*

A conventional repeater, without knowledge of TDD DL/UL configuration of the system, may turn on two amplification chains for both DL and UL directions in every slot as shown in Figure 5, regardless of whether a slot is a DL slot or an UL slot. With both amplification chains on, the transmitted signal will loop back to the receiver side via same-side coupling between two chains (the red loop) and the other-side coupling within each chain as shown in Figure 5 and may lead to unstable oscillation, if the amplification gain exceeds a maximum gain limit $G_{max}$. The gain limit $G_{max}$ depends on the coupling matrix between transmitter and receiver. Note that the same side coupling with aligned beam directions is much stronger than the other side coupling without aligned beam directions.

If the TDD DL/UL configuration can be provided to a smart repeater, the smart repeater only needs to turn on one chain based on whether the slot is a DL slot or an UL slot. In this case, there is only other-side antenna coupling, which is much weaker than the same-side antenna coupling. Therefore, a smart repeater with knowledge of TDD DL/UL configuration can operate at a much higher maximum stable gain than a conventional repeater. In our OTA experiment (Section IV), it is observed that the gain limit $G_{max}$ can be improved by up to 20dB with TDD information.

Another side information is the scheduled beam direction for AC link. For a conventional repeater, due to lack of such information, a fixed broad beam is used on the AC link to cover all potential beam directions with a smaller array gain, i.e. $f_{BF} < 1$. But for a smart repeater, if scheduled beam information can be provided by the network, the smart repeater can form a narrow beam toward the scheduled UE with the maximum antenna array gain, i.e. $f_{BF} = 1$.

Note that the TDD DL/UL configuration can be semi-static, which does not change in the time scale of scheduling slots; while the scheduled beam information can be dynamic, which changes in time scale of scheduling slots and has larger control signaling overhead. In order to understand how much performance improvement can be achieved with different levels of side information, in section III, we consider two types of smart repeaters: *semi-smart repeater* with only TDD-awareness, and *smart repeater* with both TDD-awareness and scheduled beam information.

Comparing with an amplify-forward repeater, a decode-forward relay node has larger implementation complexity and latency, because it needs additional digital components to decode the received packet and then encode and transmit to the next hop.

## III. SIMULATION RESULTS

### A. Simulation Assumptions

We consider a deployment of gNBs and relay nodes in a Manhattan grid as shown in Figure 6, where gNBs are placed at intersections along every even street, and relay nodes are placed at intersections along every odd street. Each gNB has four sectors, covering east, west, north, and south directions. Each relay node connects to one gNB, from which it receives the strongest signal via one of its two BH sectors pointing to north or south directions and provides service to the UEs along the adjacent street via its two AC sectors pointing to east and west directions. It can be seen that if relay nodes are not deployed, there will be no coverage on odd streets, except in the areas close to the intersections.

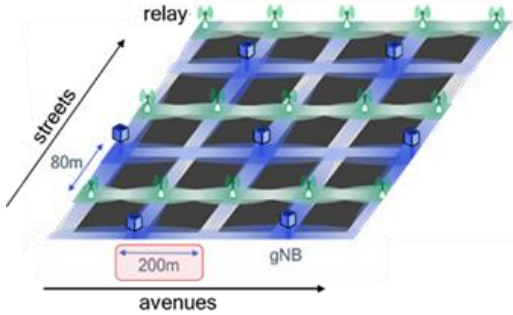

*Figure 6: Manhattan grid deployment*

We consider a simulation area of $2000 \times 2000$ meters with a total of 84 gNBs and 156 relay nodes. 840 UEs (10 UEs per gNB) are randomly dropped outdoor and along the streets and avenues. For a UE, the serving node (a gNB or a relay node) is determined as a node from which the received power at the UE is the largest. If the serving node is a gNB, the UE is called a direct UE of the gNB, otherwise (when the serving node is a relay node, which is connected to a gNB via BH link), the UE is called an indirect UE of the gNB. Detailed simulation parameters are shown in Table 1. It is assumed that there are buildings along the sides of streets and avenues and the wireless signals are diffracted by the building.

*Table 1: Simulation parameters*

| Parameter | Value |
| --- | --- |
| Topology | Manhattan grid, 84 gNBs, 156 Relays<br>inter-(avenue, street) distance = (200, 80)m;<br>(avenue, street) width = (14, 8)m |
| Antenna (Azi x Ele) | • gNB: 16 × 4 per sector<br>• Relay: 4 × 1 per BH sector; 16 × 4 per AC sector<br>• UE: 2 × 1 |
| Channel | • (fc, BW)=(28, 0.8)GHz, power per PA=7dBm<br>• Pathloss exponent<br>  o BH: 2 if distance<200m; 3.2 o.w.<br>  o AC: 2 if distance<30m; 3.2 o.w.<br>• Shadow fading: 8dB (AC link), 4dB (BH link)<br>• Knife-edge diffraction [10], reflection not modeled |
| Relay Param | $[G_{max,A}]_{dB} = 50dB$, $[G_{max,B}]_{dB} = 70dB$, $[\delta_{NF}]_{dB} = 1dB$ |
| Beamformer | Azimuth:<br>• For AC links of conventional repeaters: fixed broad beam.<br>• For other cases: constant phase offset (CPO) beam steering<br>Elevation: max array gain of CPO beams. |
| Link-association | One-hop RX power |
| Scheduler | • Round-robin;<br>• spatial reuse between direct and indirect UEs can be used for decode-forward relay scheme. |
| Inter-cell Interference | • Based on azimuth beam pattern + fixed elevation gain<br>• No interference between AC and BH links |

We designed a broad beam pattern for the conventional repeater with 16 azimuth antenna elements that fit the Manhattan grid geometry. This beam pattern has a beamforming loss factor of $[f_{BF}]_{dB} \leq -8dB$ relative to a smart repeater and a decode-forward relay node with the maximum array gain.

We consider the following cases for performance comparison:

- **noRepeaterRelay**: no repeater or relay nodes.
- **conventionalRepeater:** conventional repeater with a single broad beam and $[G_{max,A}]_{dB} = 50$dB;
- **semi-smartRepeater:** TDD-aware repeater, that uses the same broad beam as conventional repeater but with a higher amplification gain limit $[G_{max,B}]_{dB} = 70$dB;
- **smartRepeater:** TDD-aware repeater with scheduled beam information that can point its beam to a scheduled UE, with a gain limit $[G_{max,B}]_{dB} = 70$dB;
- **halfDuplexRelay-SpatialReuse, halfDuplexRelay-NoSpatialReuse:** half-duplex decode-forward relay node with and without spatial reuse scheme.
- **fullDuplexRelay-SpatialReuse**, **fullDuplexRelay-NoSpatialReuse**: full duplex decode-forward relay node with and without spatial reuse scheme. In these two cases, the self-interference for full-duplex decode-forward relay node is modeled, assuming 130dB isolation between Tx and Rx.

The scheduler follows a simple round-robin scheme, where for each sector of a gNB, the gNB selects and serves UEs out of all its associated direct or indirect UEs one-by-one over consecutive slots. An advanced scheduler, with *spatial reuse*, can be enabled for cases with decode-forward relay nodes as shown in Figure 7. When an indirect UE is scheduled by a gNB via a decode-forward relay node in a slot, there may be a time period within the slot that the BH link is not used in parallel with the AC link for the indirect UE, e.g. for the case with half-duplex operation, or for the case with full-duplex operation when the BH link with larger rate finishes its TX earlier than the AC link. During that time period, the gNB can schedule another direct UE while the relay node still serves the indirect UE on the AC link.

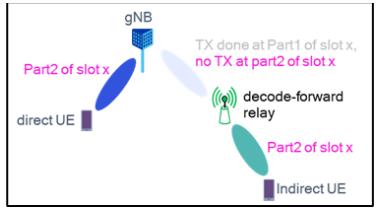

*Figure 7: Example of spatial reuse for decode-forward relay node*

For the cases of conventional repeater or semi-smart repeater, the repeater is assumed to be always on due to lack of dynamic scheduling information, which will always generate interference to other links. For the cases of smart repeater or a decode-forward relay node, the repeater/relay node can be turned on or off based on whether an associated indirect UE is scheduled by the gNB, to reduce interference to other links.

### B. Simulation Results

Under the simulation assumptions shown in Section III.A, we evaluate the DL performance for various cases and the distributions (CDF) of various metrics are shown in Figure 8-Figure 10. Note the number of indirect UEs connected via repeater/relay nodes are different for different cases, the percentage of indirect UEs are (32%, 38.5%, 45.7%) respectively for cases with (conventional repeaters, semi-smart repeaters, smart repeaters as well as decode-forward relays).

Figure 8 presents the CDF of end-to-end effective SINR of direct and indirect UEs. While all types of relay nodes can improve the SINR performance compared to the case without relay nodes, smart repeaters and full-duplex decode-and-forward relays offer the best performance.

Figure 9 shows the CDFs of achieved rates over scheduled slots for all indirect UEs. *smartRepeater*'s performance is only slightly worse than the *fullDuplexRelay*, and much better than the other solutions. It can further be observed that the advanced scheduler schemes, with spatial reuse, has not much impact on the indirect UE's performance. They may indeed slightly worsen the performance, due to more interference caused by the direct UEs.

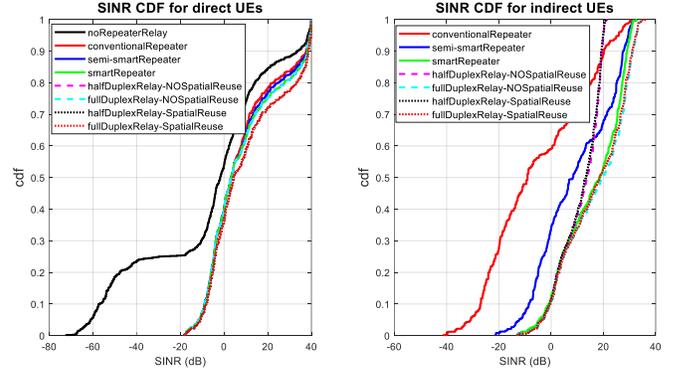

*Figure 8: CDF of effective DL SINR for direct and indirect UEs*

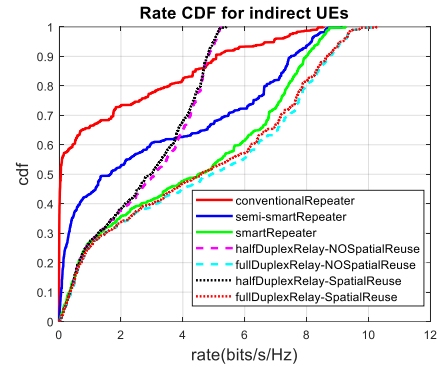

*Figure 9: CDF of DL spectrum efficiency for indirect UEs*

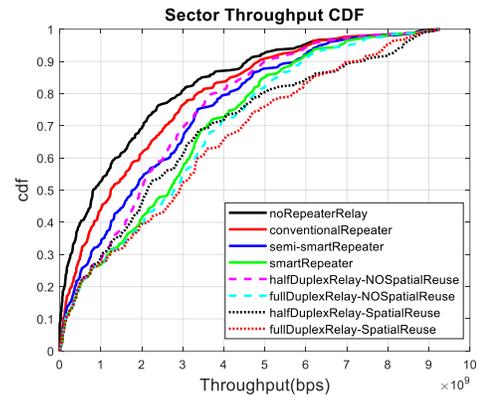

*Figure 10: CDF of DL sector throughput*

Figure 10 shows the CDFs of sector throughputs including both direct and indirect UEs. As expected, *smartRepeater* outperforms the other analog-and-forward repeater solutions. It further provides a better overall sector throughput compared to the half-duplex decode-and-forward relay with no spatial reuse, and almost a similar performance to that of the full-duplex decode-and-forward relay with no spatial reuse. Advanced scheduling schemes, employing spatial reuse, allow for better resource utilization at the network side and hence an improved sector throughput for both half-duplex and full-duplex decode-and-forward relays.

## IV. OVER-THE-AIR TESTS

To validate the analytical and simulation results presented in the previous sections, over-the-air tests, at 28GHz carrier frequency, were conducted using repeaters (with various level of available side information), in both indoor and outdoor environments (Figure 11). The repeater has two units for the BH and AC links respectively, where each unit comprises two arrays of $16 \times 8$ antenna elements. To test the conventional repeaters (without TDD information), the two arrays on each side are simultaneously active to forward signals in both UL and DL directions. For TDD-aware repeaters, only one array per unit is active in each slot, depending on the TDD DL/UL state of the slot.

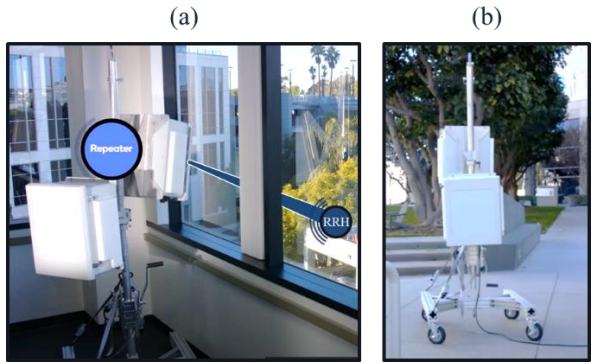

Figure 11: OTA test setup: (a) outdoor-to-indoor coverage extension, and (b) outdoor.

The repeater stability test reveals that TDD-awareness allows driving the repeater's maximum amplification gain at a higher level before repeaters entering an instable region. The maximum gain depends on the environment (e.g., the distribution of near-by clutters) and repeater's beamforming configuration. Compared to the conventional repeaters (without TDD information), TDD-aware repeaters can provide up to 10dB and 20dB higher gains in indoor and outdoor environments respectively. This, in turn, results in up to 100% improvement in DL throughput in the indoor test setup.

Without AC link beamforming information, a fixed broad beam of 30-degree beam-width is used. On the other hand, narrow beams of 6-degree beam-width are used, when the AC link beam info is available. Adaptive AC link beamforming offers up to 7dB extra array gain compared to the solutions with fixed AC beam.

All types of repeaters demonstrate great improvements in the DL throughput and range, compared to the scenarios without repeaters. In outdoor test setups, smart repeaters outperform conventional repeaters with up to 40% higher offered DL throughput.

## V. CONCLUSIONS

In this paper, we analyzed and compared the performance of different types of repeater/relay nodes: conventional amplify-forward repeater, semi-smart/smart amplify-forward repeaters with different levels of side information, and half-duplex/full-duplex decode-forward relay node with and without spatial reuse for multi-user scheduling. The practical constraints and factors, e.g. the maximum amplification gain for stability of amplify-forward repeaters, the noise figure difference between amplify-forward repeater and decode-forward relay node, self-interference and inter-cell interference, etc., are considered in the analysis and/or simulation.

The end-to-end effective SINRs and achievable rates for various repeater/relay nodes were analytically derived and extended to a general N-hop network. The system-level studies were also conducted using a Manhattan grid deployment. The performance of different types of repeaters was further evaluated for a repeater prototype implementation in 28 GHz and using different OTA test setups. We discussed performance and complexity tradeoff between these types of relay nodes. The conventional repeaters have the least complexity and latency but achieve the lowest data rate, the full-duplex decode-forward relays can achieve the highest throughput at the cost of higher complexity and latency. For the half-duplex decode-forward relays, the half-duplex penalty can be compensated partly via spatial reuse in multi-user scheduling. The smart amplify-forward repeaters with different levels of side information strike a good balance between performance and complexity.

## ACKNOWLEGMENT

The authors would like to thank Juergen Cezanne (Senior Director, Technology) of Qualcomm Technologies, Inc. for many fruitful discussions, especially on beam design, as well as the Qualcomm millimeter wave prototyping team for developing a repeater implementation and conducting the OTA tests.

## REFERENCES

[1] 3GPP TR 38.874 "Study on Integrated Access and Backhaul (Release-16)", V16.0.0, Dec. 2018.

[2] 3GPP TS 38.106, "NR Repeater Radio Transmission and Reception (Release-17)", V17.6.0, Sept. 2023.

[3] 3GPP TR 38.867, "Study on NR Network-Contolled Repeaters (Release 18)", V18.0.0, Sept. 2022.

[4] 3GPP TS 23.304 "Proximity based Services (ProSe) in the 5G System (5GS) (Release-18)", V18.3.0, Sept. 2023.

[5] G. Liu, F. R. Yu, H. Ji, V. C. M. Leung, and X. Li, ''In-band full-duplex relaying: A survey, research issues and challenges,'' IEEE Commun. Surveys Tuts., vol. 17, no. 2, pp. 500–524, 2nd Quart., 2015.

[6] J. N. Laneman, D. N. C. Tse, and G. W. Wornell, "Cooperative diversity in wireless networks: Efficient protocols and outage behavior," IEEE Trans. Inform. Theory, vol. 50, no. 12, pp. 3062–3080, Dec. 2004

[7] M. R. Souryal and B. R. Vojcic, "Performance of amplify-andforward and decode-and-forward relaying in Rayleigh fading with turbo codes," in Proc. ICASSP 2006, vol. 4, pp. 681-684, 2006.

[8] G. Levin and S. Loyka, "Amplify-and-Forward Versus Decode-and-Forward Relaying: Which is Better?", International Zurich Seminar on Communications (IZS), February 29 – March 2, 2012

[9] J. Xin, S. Xu, S. Xiong, H. Xu, H. Zhang, "A Survey on Network Controlled Repeater Technology", IEEE the 8[th] International Conference on Computer and Communications, Dec. 9-12 2022.

[10] Radio and Microwave Wireless Systems", University of Toronto ECE422 Course lecture note, Prof. Sean V. Hum.